\documentclass[twocolumn,apj,iop,numberedappendix]{emulateapj}
\usepackage{amsmath}
\usepackage{graphicx}
\usepackage[hyperfootnotes=false,colorlinks,citecolor=blue,linkcolor=blue,urlcolor=blue]{hyperref}
\DeclareMathAlphabet{\mathbfit}{OT1}{cmr}{bx}{it}
\renewcommand{\vec}[1]{\mathbfit{#1}}
\newcommand{\mat}[1]{\mathbfit{#1}}
\newcommand{\Mo}{\mathrm{M}_{\odot}}

\begin{document}

\title{On determining the shape of matter distributions}
\shorttitle{On determining the shape of matter distributions}

\author{Marcel Zemp$^{1}$, Oleg Y. Gnedin$^{1}$, Nickolay Y. Gnedin$^{2,3,4}$ and Andrey V. Kravtsov$^{3,4}$}
\affil{$^{1}$ Department of Astronomy, University of Michigan, Ann Arbor, MI 48109, USA\\
$^{2}$ Particle Astrophysics Center, Fermi National Accelerator Laboratory, Batavia, IL 60510, USA\\
$^{3}$ Kavli Institute for Cosmological Physics, University of Chicago, Chicago, IL 60637, USA\\
$^{4}$ Department of Astronomy \& Astrophysics, University of Chicago, Chicago, IL 60637 USA}

\email{mzemp@umich.edu}

\shortauthors{Zemp et al.}

\begin{abstract}
A basic property of objects, like galaxies and halos that form in cosmological structure formation simulations, is their shape.
Here, we critically investigate shape determination methods that are commonly used in the literature.
It is found that using an enclosed integration volume and weight factors $r^{-2}$ and $r_\mathrm{ell}^{-2}$ (elliptical radius) for the contribution of each particle or volume element in the shape tensor leads to biased axis ratios and smoothing of details when calculating the local shape as a function of distance from the center.
To determine the local shape of matter distributions as a function of distance for well resolved objects (typically more than $\mathcal{O}(10^4)$ particles), we advocate a method that (1) uses an ellipsoidal shell (homoeoid) as an integration volume without any weight factors in the shape tensor and (2) removes subhalos.
\end{abstract}

\keywords{methods: data analysis --- methods: numerical}

\section{Introduction}

Typically, the distribution of matter in objects that form in cosmological structure formation simulations is crudely described by spherically averaged density profiles \cite[e.g.][]{1996ApJ...462..563N,1998ApJ...499L...5M}.
But real halos are not spherically symmetric and a natural extension is to describe the iso-density contours as surfaces of ellipsoids.
There is a wealth of literature with many different methods that are used to measure the local shape of a mass distribution \citep[e.g.][]{1983MNRAS.202.1159G,1988ApJ...327..507F,1991ApJ...368..325K,1991ApJ...378..496D,1992ApJ...399..405W,1996MNRAS.281..716C,2002ApJ...574..538J,2004IAUS..220..421S,2004ApJ...611L..73K,2004ApJ...616...27B,2005ApJ...627..647B,2006MNRAS.367.1781A,2007ApJ...671.1135K,2007MNRAS.376..215B,2007MNRAS.377...50H,2008ApJ...681.1076D,2008MNRAS.385.1859W,2010ApJ...720L..62K,2010MNRAS.405.1119K,2011ApJ...734...93L,2011arXiv1104.1566V}.
Their common goal is to recover the iso-density surfaces of the underlying matter distribution.
Other characteristics, such as the potential, can also be used to describe the objects \citep[e.g.][]{2004IAUS..220..421S,2007MNRAS.377...50H,2010ApJ...720L..62K,2011ApJ...734...93L}.

Unfortunately, the literature lacks a systematic comparison of the different methods - especially under controlled conditions where the exact shape is known.
For some notable exceptions, see e.g. \cite{2006MNRAS.367.1781A,2004ApJ...611L..73K}.
But as far as we know, there is no publication that investigates the different methods under controlled conditions with known shape as it is done in this paper.
Presumably, many of the quantitative discrepancies in the literature originate in the various methods that are used for determining the shape.
This work is intended to shed some light on the effects and systematics of the various methods that are based on an iterative procedure that uses a shape tensor with different weighting schemes and integration volumes.
The influence of the local mass density profile on the capability of the shape finding method to recover the iso-density contours is also investigated.

\section{Background}\label{sec:bg}

From classical mechanics, the relation between the angular momentum vector $\vec{L}$ and the angular velocity vector $\boldsymbol\omega$ of a body is given by 
\begin{equation}
\vec{L} = \mat{I} \boldsymbol\omega~,
\end{equation}
where $\mat{I}$ is the moment of inertia tensor defined by
\begin{equation}\label{eq:mit}
\mat{I} \equiv \int_V \rho(\vec{r}) (\vec{r}^2 \mat{1} - \vec{r} \vec{r}^T) \mathrm{d} V~,
\end{equation}
where the integration is over the whole volume of the body and $\mat{1}$ is the identity tensor.
Here, $\rho(\vec{r})$ is the mass density at the location of the volume element d$V$ pointed by the position vector $\vec{r}$ with respect to the center of the mass distribution.
By defining the tensor
\begin{equation}\label{eq:M}
\mat{M} \equiv \int_V \rho(\vec{r}) \vec{r} \vec{r}^T \mathrm{d} V~,
\end{equation}
which is the second moment of the mass distribution, it follows that
\begin{equation}
\mat{I} = \mathrm{tr}(\mat{M}) \mat{1} - \mat{M}~.
\end{equation}
Hence, the tensor $\mat{M}$ is the fundamental quantity that describes how the matter is distributed.

We now define the shape tensor as
\begin{equation}\label{eq:shapetensor}
\mat{S} \equiv \frac{\mat{M}}{M_\mathrm{tot}} = \frac{\int_V \rho(\vec{r}) \vec{r} \vec{r}^T \mathrm{d} V}{\int_V \rho(\vec{r}) \mathrm{d} V}
\end{equation}
where
\begin{equation}
M_\mathrm{tot} = \int_V \rho(\vec{r}) \mathrm{d} V
\end{equation}
is the total mass of the body.
The shape tensor has units of length squared.
For a discrete set of particles with
\begin{equation}
\rho(\vec{r}) = \sum_k m_k \delta(\vec{r} - \vec{r}_k)~,
\end{equation}
we obtain for the individual elements of the shape tensor
\begin{equation}
S_{ij} = \frac{\sum_k m_k (\vec{r}_k)_i (\vec{r}_k)_j}{\sum_k m_k}
\end{equation}
where $(\vec{r}_k)_j$ denotes the $j$ component of the position vector of the $k$-th particle and the summation is over all particles within the integration volume $V$.
The tensors $\mat{S}$ and $\mat{M}$ describe how the mass is distributed, hence our choice for naming $\mat{S}$ the shape tensor.

The tensors $\mat{I}$ and $\mat{M}$ have the same eigenvectors.
If $m$ is an eigenvalue of $\mat{M}$, then $\mathrm{tr}(\mat{M}) - m$ is an eigenvalue of $\mat{I}$.
The detailed meaning of the eigenvalues depends on the integration volume and the mass distribution (i.e. density profile).
For example, for a thin ellipsoidal shell (a thin homoeoid) of uniform density, the eigenvalues of $\mat{M}$ are $M_\mathrm{ES}a^2/3, M_\mathrm{ES}b^2/3$ and $M_\mathrm{ES}c^2/3$ ($M_\mathrm{ES}$ is the mass in the ellipsoidal shell).
Whereas for an ellipsoid of uniform density the eigenvalues are $M_\mathrm{E}a^2/5, M_\mathrm{E}b^2/5$ and $M_\mathrm{E}c^2/5$ ($M_\mathrm{E}$ is the mass in the ellipsoid).

Unfortunately, the tensor $\mat{M}$ (Equation (\ref{eq:M})) is often inaccurately denoted as the moment of inertia tensor in the astronomy and astrophysics literature.
This probably goes back to \cite{1987gady.book.....B} (Page 494, Equation 8-11), where $\mat{M}$ was called the moment of inertia tensor.
Fortunately, this was corrected in the second edition \citep[page 796, Equation D-39]{2008gady.book.....B}.

\section{Methods}\label{sec:methods}

The shape tensor can be generalized by using an additional weight function $w(\vec{r})$
\begin{equation}\label{eq:shapetensorgeneral}
\mat{S} = \frac{\int_V \rho(\vec{r}) w(\vec{r}) \vec{r} \vec{r}^T \mathrm{d} V}{\int_V \rho(\vec{r}) \mathrm{d} V}~.
\end{equation}
By setting $w(\vec{r}) = 1$ and choosing $\rho(\vec{r})$ to be the mass density we obtain our standard definition (Equation (\ref{eq:shapetensor})).
Other choices are also possible.
For example, a weighting by number with $\rho(\vec{r}) = \sum_k \delta(\vec{r} - \vec{r}_k)$ being the number density (which is, of course, equivalent to the mass density weighting if all the particles have equal mass).
Or $\rho(\vec{r}) = \sum_k \delta(\vec{r} - \vec{r}_k)/\rho_k$ where $\rho_k$ is the local density of the particle like in \cite{2008MNRAS.385.1859W}.
If one is interested in the shape of a matter distribution where the particles or volume elements can have a different mass (e.g. for gas and stars), it is essential to use $\rho(\vec{r})$ as the mass density.
Here, we only use $\rho(\vec{r})$ as the mass density.
Throughout the paper, we use the elliptical radius $r_\mathrm{ell}$ for distances from the center for ellipsoidal shapes.
The elliptical radius $r_\mathrm{ell}$ (see also Equation (\ref{eq:rell})) is the semi-major axis of the local homoeoid or ellipsoid.

\begin{deluxetable}{ccc}[t]
	\tablecolumns{3}
	\tablewidth{0pt}
	\tablecaption{Summary of methods \label{tab:methodsummary}}
	\tablehead{\colhead{Method} & \colhead{$w(\vec{r})$} & \colhead{$V$}}
	\startdata
	S1 & 1 & ellipsoidal shell\\
	S2 & $r^{-2}$ & ellipsoidal shell\\
	S3 & $r_\mathrm{ell}^{-2}$ & ellipsoidal shell\\
	E1 & 1 & enclosed ellipsoid\\
	E2 & $r^{-2}$ & enclosed ellipsoid\\
	E3 & $r_\mathrm{ell}^{-2}$ & enclosed ellipsoid
	\enddata
\end{deluxetable}

We concentrate on 6 different methods for determining the shape of a matter distribution (see also Table \ref{tab:methodsummary}).
These methods differ by using a different integration volume $V$ and different weight functions $w(\vec{r})$.
For calculating the \textit{local} shape at a distance $r_\mathrm{ell}$, in the methods with a starting letter S, the integration is over an ellipsoidal shell (homoeoid) volume centered at $r_\mathrm{ell}$ (in logarithmic space).
In the methods with first letter E, the integration is over the whole enclosed ellipsoidal volume within $r_\mathrm{ell}$.
For the different weight functions $w(\vec{r})$, we use (1) $w(\vec{r}) = 1$, (2) $w(\vec{r}) = r^{-2}$ and (3) $w(\vec{r}) = r_\mathrm{ell}^{-2}$.
The elliptical radius is given by
\begin{equation}\label{eq:rell}
r_\mathrm{ell} = \sqrt{x_\mathrm{ell}^2 + \frac{y_\mathrm{ell}^2}{(b/a)^2} + \frac{z_\mathrm{ell}^2}{(c/a)^2}}
\end{equation}
where $(x_\mathrm{ell},y_\mathrm{ell},z_\mathrm{ell})$ are the coordinates of the volume element or particle in the eigenvector coordinate system of the ellipsoid, i.e. $r_\mathrm{ell}$ corresponds to the semi-major axis $a$ of the ellipsoid surface through that particle or volume element.
Additionally, we also check for the importance of the removal of subhalos.
Cases where we removed the subhalos are marked with a --, cases where they remained by a +.

In order to calculate the local shape at a distance $r_\mathrm{ell}$ from the center, we use an iteration method \citep[e.g.][]{1991ApJ...368..325K,1991ApJ...378..496D,1992ApJ...399..405W} and start with a spherically symmetric integration volume (shell or sphere).
Then the shape tensor is calculated according to the different methods.
By diagonalizing $\mat{S}$ we get the eigenvectors and eigenvalues at distance $r_\mathrm{ell}$.
The eigenvectors give the directions of the semi-principal axes.
The eigenvalues of $\mat{S}$ for the method S1 are $a^2/3$, $b^2/3$ and $c^2/3$ where $a$, $b$ and $c$ are the semi-principal axes with $a \geq b \geq c$ -- at least in the thin homoeoid approximation where the density is uniform.
Hence, the square roots of the eigenvalues are proportional to the lengths of the semi-principal axes for method S1 and we can readily calculate the axis ratios $b/a$ and $c/a$.
For method S3 we expect to get the same axis ratios as for method S1 since dividing by the semi-major axis $a=r_\mathrm{ell}$ squared, which is a constant for a thin ellipsoidal shell, just changes the geometrical meaning and normalization of the eigenvalues but not the axis ratios.

For the other methods it is not clear what the detailed geometrical meaning of the eigenvalues is.
For the methods that use the enclosed ellipsoidal volume, this will also depend on the mass density profile.
The $r^{-2}$ weighting projects the volume elements d$V$ onto the unit sphere.
This projection complicates the physical interpretation of this method.
It is generally assumed though that the eigenvalues of $\mat{S}$ in these cases are still proportional to the semi-major axes squared.
Hence, we calculate the axis ratio for the other methods the same way as for methods S1 and S3 -- as it is generally done in the literature.

We then keep the length of the semi-major axis fixed (but the orientation can change) and calculate $\mat{S}$ again by summing over all particles within the new deformed integration volume (homoeoid or ellipsoid) with semi-major axis $a=r_\mathrm{ell}$ and axis ratios $b/a$ and $c/a$ but with the new orientation.
For the shape determination we allow volume elements or particles to be in several bins/shells.
Of course this is naturally the case when using an enclosed ellipsoidal volume.
It is also necessary when using an ellipsoidal shell since neighboring shells can overlap due to slightly different orientation and axis ratios.
This iteration is repeated until convergence is reached.
As a convergence criterion we require that the fractional difference between two iteration steps in both axis ratios is smaller than $10^{-3}$.

For methods using the shape tensor, it is important to use an iteration method that allows the algorithm to adapt the integration volume to the a priori unknown shape of the object.
Often one also finds in the literature that no iteration procedure is used and just a simple spherical shell or enclosed sphere is used as the integration volume in order to calculate the shape \citep[e.g.][]{1983MNRAS.202.1159G,1988ApJ...327..507F,1996MNRAS.281..716C,2004ApJ...616...27B,2005ApJ...627..647B,2010MNRAS.405.1119K}.
To us the physical meaning of the outcome of such a procedure is unclear and we do not further pursue it here.

A further method for calculating the shape of contours is by selecting particles by their local density \citep[e.g.][]{2002ApJ...574..538J,2008MNRAS.385.1859W,2011arXiv1104.1566V} or potential \citep[e.g.][]{2004IAUS..220..421S,2007MNRAS.377...50H,2010ApJ...720L..62K}.
There, no iteration procedure is needed.

Often one also finds in the literature, that the moment of inertia tensor $\mat{I}$ (Equation (\ref{eq:mit})) in combination with an enclosed ellipsoidal integration volume is used for calculating the axis ratios.
This procedure assumes relations between the eigenvalues and semi-principle axes that are strictly valid only for a uniform ellipsoid or homoeoid \cite[e.g.][]{2007MNRAS.376..215B,2008MNRAS.385.1859W}.
For a thin homoeoid this is fine (under that assumption that the local density is constant in the shell) but for the enclosed ellipsoidal integration volume, the result is made equivalent to the method that just uses the shape tensor by construction.

\section{Controlled conditions}\label{sec:cc}

First, we examine the behavior of the different methods under controlled conditions where we know the correct shape.
For this purpose, we set up various model halos that have different density, shape and orientation profiles with \textsc{halogen} \citep{2008MNRAS.386.1543Z}.

\subsection{Models}

\textsc{halogen} can generate random realizations of spherical halos with $\alpha\beta\gamma$-profiles \citep{1996MNRAS.278..488Z}
\begin{equation}\label{eq:abc}
\rho(r) = \frac{\rho_0}{(r/r_\mathrm{s})^\gamma [1 + (r/r_\mathrm{s})^\alpha]^{[(\beta-\gamma)/\alpha]}}
\end{equation}
in equilibrium, where an importance sampling method (multimass technique) can be applied.
Here, we just interpret the spherical radius $r$ in the $\alpha\beta\gamma$-profiles as semi-major axis $a=r_\mathrm{ell}$ of a surface of an ellipsoid.
For the generation of a uniform distribution of points on a surface of an arbitrary shaped ellipsoid, which is needed for setting up an ellipsoid with a given density profile, a method as outlined in Section 2.5.5 of \cite{2007smcm.book.....R} is used.
Since we only care about the spatial distribution of the matter for our purpose, no velocities are assigned to the sampled particles.

For the variation of the axis ratios and orientation with distance, we use a simple parametrization of the form
\begin{equation}\label{eq:vp}
x(r_\mathrm{ell}) = s_x \log_{10}(r_\mathrm{ell}/r_{0,x}) + x_0~.
\end{equation}
Here, $x$ can be $b/a$, $c/a$, $\theta_1$, $\theta_2$ and $\theta_3$, respectively.
The angles $\theta_1$, $\theta_2$ and $\theta_3$ are the Euler angles of an active $z-x^\prime-z^{\prime\prime}$ rotation.
This allows us to twist the orientation of the principal axes as a function of distance.

We use a generalized NFW \citep{1996ApJ...462..563N} form for the density profile of the halos, i.e. we set $\alpha=1$, $\beta=3$, and use 3 different values for the inner slope, i.e. $\gamma$ = 0, 1 and 2.
The ellipsoidal halos are sampled with $10^7$ particles of the same mass within 10 $r_\mathrm{s}$ (no multimass technique applied).
For some cases also different resolution halos with up to $10^8$ particles within 10 $r_\mathrm{s}$ are used.
To compare to current state-of-the-art cosmological structure formation simulations: hydrodynamical simulations have reached $\mathcal{O}(10^7)$ particles per halo \citep[e.g.][]{2011arXiv1103.6030G,2011ZempBaryonicImpact} whereas halos in dissipationless N-body simulations are even resolved with $\mathcal{O}(10^9)$ particles \citep[e.g.][]{2008MNRAS.391.1685S,2009MNRAS.398L..21S}.
Beyond 10 $r_\mathrm{s}$ an exponential cut-off of the mass density profile is applied in order to keep the total mass finite (for more details see \cite{2008MNRAS.386.1543Z}).
With a resolution of $10^7$ particles, one can roughly sample an NFW profile down to 0.1 $r_\mathrm{s}$.
The resolved scale depends on the inner slope $\gamma$.
For $\gamma = 2$, this scale is smaller and for $\gamma = 0$ it is larger (for more details see \cite{2008MNRAS.386.1543Z}).
Hence, for all profiles in the following plots only the range 0.1--10 $r_\mathrm{s}$ is shown.

\subsection{Constant axis ratios - aligned orientation}\label{sec:constant_aligned}

\begin{figure}
	\includegraphics[width=\columnwidth]{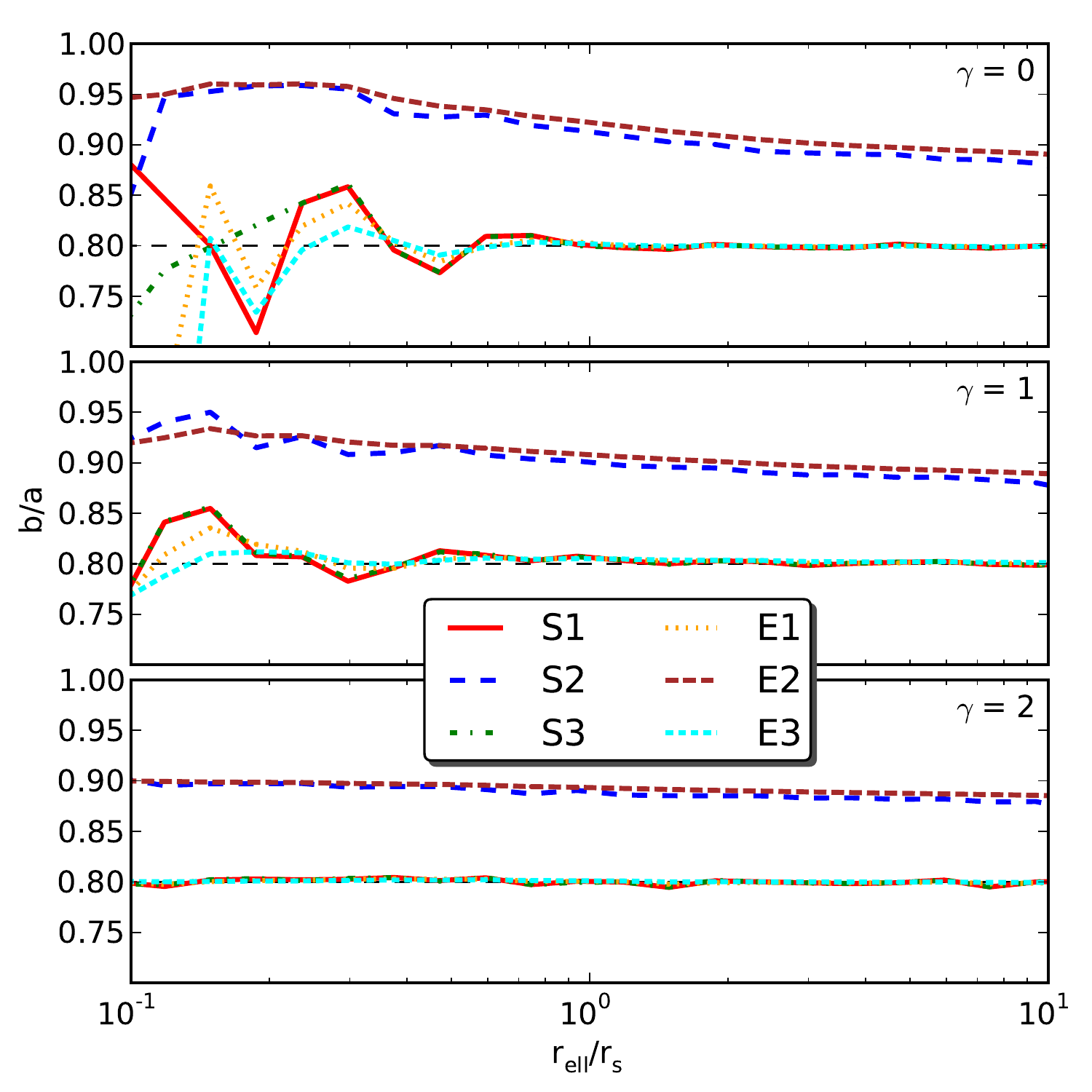}
	\caption{Measured axis ratio $b/a$ as a function of distance for halos with different inner slope $\gamma$ = 0 (top panel), 1 (middle panel) and 2 (bottom panel).
	The halos were initialized with a constant axis ratio of $b/a=0.8$ (thin dashed line).
	Except for the methods S2 and E2, all the other methods find the expected value.
	The fluctuations in the center seen for the inner slopes $\gamma$ = 0 and 1 are mainly due to resolution and decrease when increasing the sampling resolution and using a finer binning.}
	\label{fig:constant_axis_ratio_aligned_orientation_ba}
\end{figure}

As a first deviation from perfect spherical symmetry, we set up halos with constant axis ratios, while the principal axes are kept aligned at all distances.
Figure \ref{fig:constant_axis_ratio_aligned_orientation_ba} shows how the 6 different methods described in Section \ref{sec:methods} perform for our 3 halos with $\gamma$ = 0, 1 and 2.
In these models we set $b/a = 0.8$ and $c/a=0.6$.
For clarity we only show the results for $b/a$.
The findings are similar for $c/a$.

The results for the  methods S1, S3, E1 and E3 agree very well with the expected value.
The small fluctuations seen in the center are due to resolution and depend on the mass profile as well.
The fluctuations get smaller when sampling the same halo with more particles and using a finer binning.
The default binning used in this work is 10 bins dex$^{-1}$.
The number of particles in the inner most ellipsoidal shell at 0.1 $r_\mathrm{s}$ for this binning scheme ranges form ca. 2500 ($\gamma=0$) to around  $10^5$ ($\gamma=2$).
In the outer regions we have typically $\mathcal{O}(10^6)$ particles in the ellipsoidal shells.
At a given resolution, the fluctuations are larger in regions with a flat profile ($\gamma=0$) than in regions with a steep profile ($\gamma=2$).
They decrease as well in the outer regions where the profile is even steeper.
Of course, it is expected to some degree that the shape finding algorithms will have difficulty in resolving the small density contrasts from shell to shell in a nearly homogeneous region ($\gamma=0$), which explains the central fluctuations seen in this case.
Using the $r^{-2}$ weighting in methods S2 and E2 leads to a significant shift of the axis ratio towards higher values than expected.

All methods find the correct orientation of the principal axes within the well resolved range.
For example for method S1, the median deviation of $|\cos(\delta_a)-1|$, where $\delta_a$ is the angle between the measured and the correct direction of the semi-major axis $a$, is $\mathcal{O}(10^{-5})$ for all three different profile types.
For the other methods, the alignment is of comparable quality.

\subsection{Changing axis ratios - aligned orientation}\label{sec:changing_aligned}

\begin{figure}
	\includegraphics[width=\columnwidth]{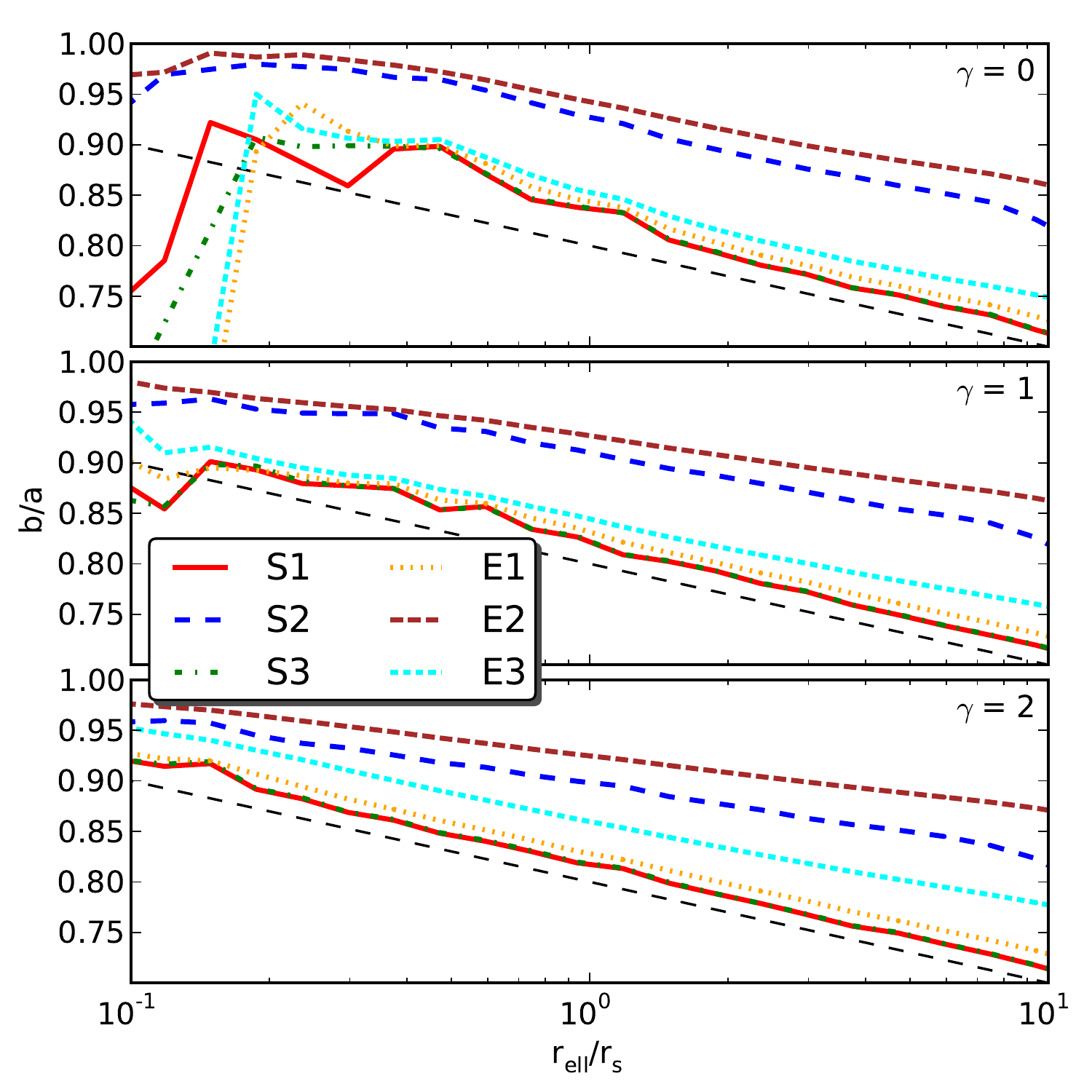}
	\caption{Measured axis ratio $b/a$ as a function of distance for halos with different inner slope $\gamma$ = 0 (top panel), 1 (middle panel) and 2 (bottom panel).
	The halos were initialized with a changing axis ratio $b/a$ as a function of distance (thin dashed line).
	We fixed $b/a=0.8$ at $r_\mathrm{s}$ and used a slope of $s_{b/a} = -0.1~\mathrm{dex}^{-1}$.
	Methods S1 and S3 give the best results, whereas methods E1 and E3 start to show systematic deviations.
	Methods S2 and E2 give again too high axis ratios.}
	\label{fig:changing_axis_ratio_aligned_orientation_ba}
\end{figure}

Of course, real halos do not have a constant axis ratio as a function of distance.
Therefore, we varied the axis ratios according to the simple parametrization given in Equation (\ref{eq:vp}).
The axis ratios were fixed at $r_\mathrm{s}$ to $b/a=0.8$ and $c/a=0.6$ and the slopes of $s_{b/a} = -0.1~\mathrm{dex}^{-1}$ and $s_{c/a} = -0.15~\mathrm{dex}^{-1}$ were used.
The condition $b/a \geq c/a$ was assured by capping the parametrization with minima and maxima.
This is not a problem within our range of interest between 0.1--10 $r_\mathrm{s}$.
The orientation of the principal axes is kept aligned with distance.

Figure \ref{fig:changing_axis_ratio_aligned_orientation_ba} shows again only the axis ratio $b/a$ as a function of distance.
Methods S1 and S3 still give the best results.
The weighting by $r^{-2}$ introduces a bias towards higher values.
Now, the methods using an enclosed volume (E1 and E3) start to show deviations as well.
This is due to the enclosed integration volume picking up information from inner regions of the halo, which has a different shape.
This leads to a lag in distance until the axis ratios can adapt.
For example, these deviations for methods E1 and E3 become larger if we choose the axis ratio to change faster as a function of distance, e.g. as for the axis ratio $c/a$ with $s_{c/a} = -0.15~\mathrm{dex}^{-1}$.

In the case shown in Figure \ref{fig:changing_axis_ratio_aligned_orientation_ba}, the axis ratio decreases with distance which leads to too high values for methods E1 and E3.
If we choose the axis ratio to increase with distance, then the methods E1 and E3 are giving too low values. 

Even the methods S1 and S3 do not perfectly reproduce the expected values.
Similar as in the case for methods E1 and E3, they lie above/below the expected value if the slope of the axis ratio is an decreasing/increasing function of distance.
The deviations for methods S1 and S3 are smaller than for methods E1 and E3.
These systematic deviations seen for methods S1 and S3 are mainly due to the local mass density profile.
In regions with a flat local profile, the systematic offset is bigger than in regions with a steep mass density profile.
Increasing the resolution and using a finer binning (i.e. smaller averaging volume) only marginally decreases the offset.
For regions with a local mass density slope $\gamma \approx 1-2$, the systematic deviations in the case of a varying axis ratio are of the order of $\mathcal{O}(0.01)$ for axis ratios for methods S1 and S3.

Again, all methods find the correct orientation of the principal axes.
The directional deviations are very small and similar to what we found in Section \ref{sec:constant_aligned}.

\subsection{Changing axis ratios - changing orientation}

\begin{figure}
	\includegraphics[width=\columnwidth]{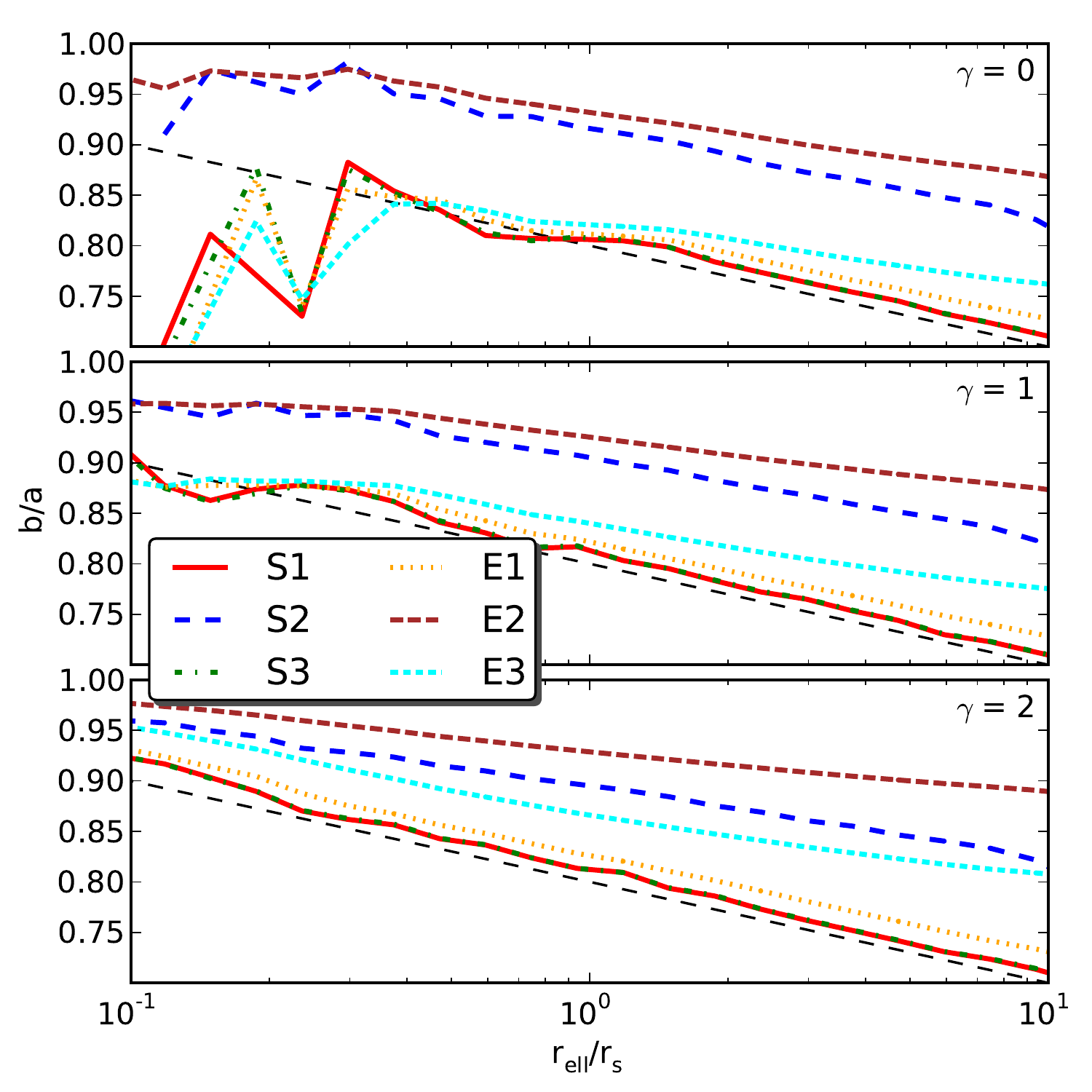}
	\caption{Measured axis ratio $b/a$ as a function of distance for halos with different inner slope $\gamma$ = 0 (top panel), 1 (middle panel) and 2 (bottom panel).
	The halos were initialized with a changing axis ratio $b/a$ as in Section \ref{sec:changing_aligned} (thin dashed line) but we varied the orientation of the principal axes as a function of distance as described in the main text.
	Methods S1 and S3 give again the best results, whereas the other methods show systematic deviations.}
	\label{fig:changing_axis_ratio_changing_orientation_ba}
\end{figure}

In real halos, the orientation of the principal axis can change as a function of distance as well.
This is parametrized again by using the functional form of Equation (\ref{eq:vp}).
The axis ratios are kept changing as in Section \ref{sec:changing_aligned}.
Additionally, we vary the alignment of the principal axes by setting ($\theta_1$,$\theta_2$,$\theta_3$) = (0.375,0.125,0.25) $\tau$ at $r_\mathrm{s}$, with $\tau \equiv 2\pi$.
For the slopes we use ($s_{\theta_1}$,$s_{\theta_2}$,$s_{\theta_3}$) = (0.05,0.05,0.05) $\tau~\mathrm{dex}^{-1}$.

Figure \ref{fig:changing_axis_ratio_changing_orientation_ba} shows the axis ratio $b/a$ as a function of distance.
If the local mass density profile is well resolved, then methods S1 and S3 are closest to the correct axis ratios.
For the other methods we see some systematic deviations which depend on the details of the axis twist.

\begin{figure}
	\includegraphics[width=\columnwidth]{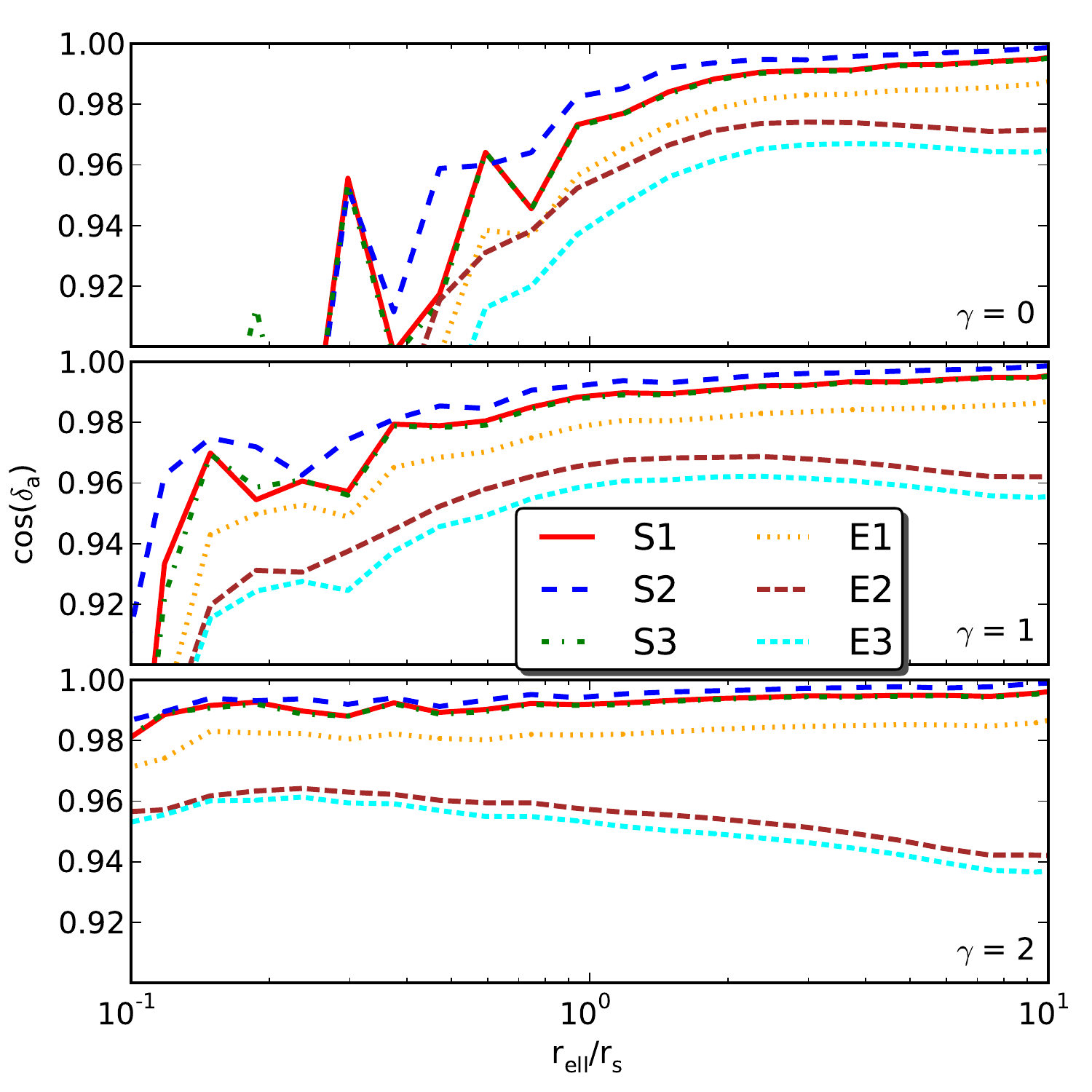}
	\caption{Cosine of the the alignment angle $\delta_a$, the angle between the measured and the correct direction of the semi-major axis $a$, as a function of distance for halos of Figure \ref{fig:changing_axis_ratio_changing_orientation_ba}.
	The methods that use an ellipsoidal shell as an integration volume are much better in recovering the local orientation of the matter distribution.}
	\label{fig:changing_axis_ratio_changing_orientation_da}
\end{figure}

The findings about the axis ratios are reflected as well in the orientation of the principal axes.
In Figure \ref{fig:changing_axis_ratio_changing_orientation_da}, we show $\cos(\delta_a)$ as a function of distance, where $\delta_a$ is the angle between the measured and the correct direction of the semi-major axis $a$.
All methods that use the enclosed integration volume show larger deviations in the orientation than the methods using a homoeoid as integration volume.
The best method for recovering the local orientation in the well resolved region in this case is S2 tightly followed by S1 and S3.
The deviations for the semi-major axis $a$, $\delta_a$, are the largest.
The deviations are smallest for the semi-minor axis $c$, i.e. we have shown the worst case in Figure \ref{fig:changing_axis_ratio_changing_orientation_da}.

\subsection{First conclusions}

We have experimented with many more mass density, shape and orientation profiles as well as different resolutions than shown here.
The findings are always the same: using an ellipsoidal shell as an integration volume without or with $r_\mathrm{ell}^{-2}$ weighting (methods S1 and S3) gives results that are closest to the expected value under controlled conditions in regions where the mass distribution is well resolved and the density contrast is high enough (i.e. no flat mass density profiles).

Methods S1 and S3 agree, since the weighting by $r_\mathrm{ell}^{-2}$ in each shell is like dividing by a different constant in each shell, which does not affect the axis ratios.
The absolute values of the eigenvalues of the shape tensor for method S3 change of course.
Hence, our preferred method is the pure form without any weighting, i.e. method S1.
All other methods lead to significant deviations that in detail depend on the mass density, shape and orientation profile.
This makes it also impossible to come up with a correction scheme that works in all cases that would allow to convert the measured axis ratios between different methods.

\section{Halos from cosmological structure formation simulations}

\begin{figure*}
	\includegraphics[width=\textwidth]{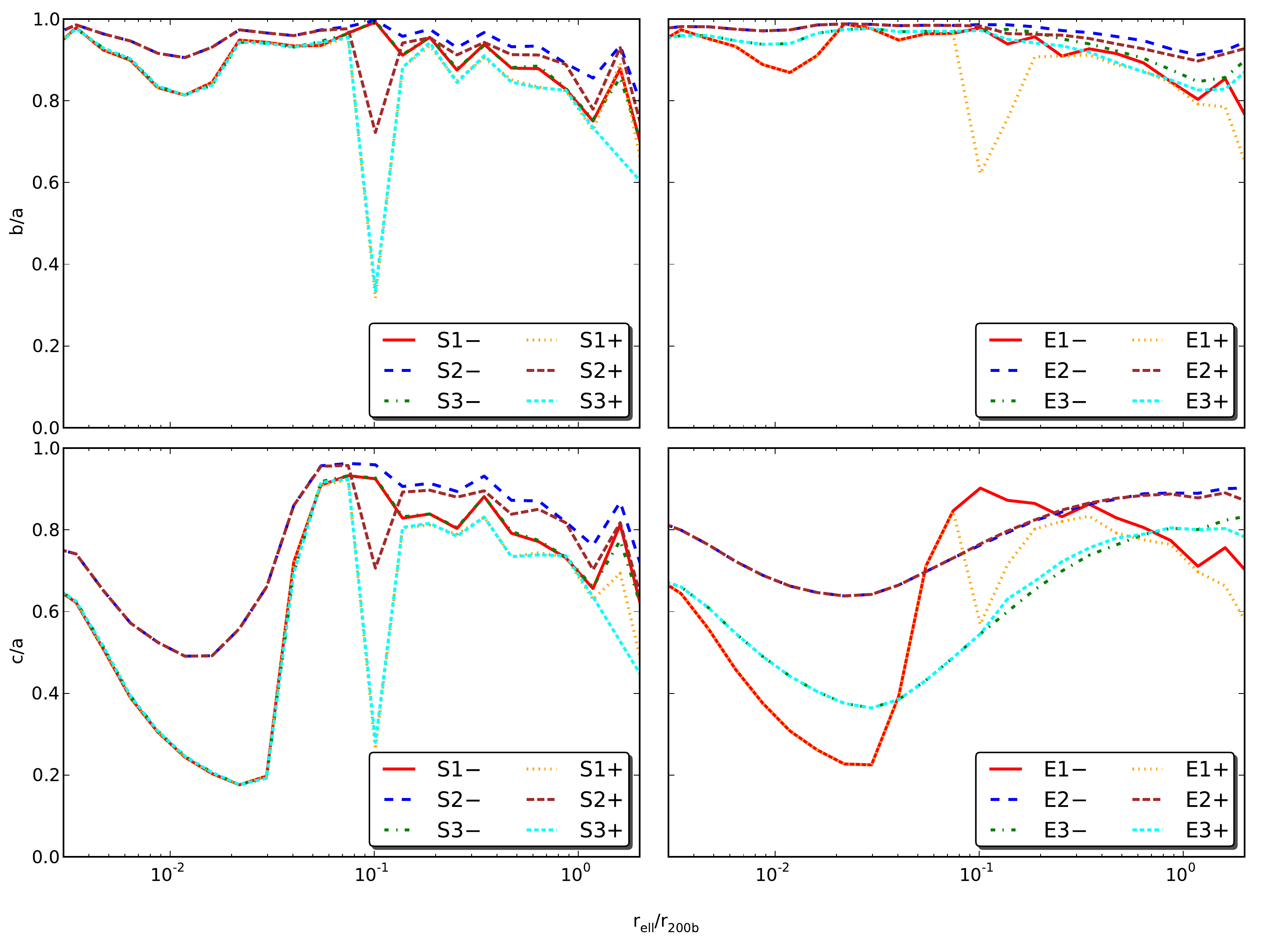}
	\caption{Measured shape of the total matter distribution of a massive halo at $z \approx 2$ in our cosmological simulation.
	In the top row we plot $b/a$ and in the bottom row $c/a$ as a function of distance.
	In the left column we show methods S1-S3 (integration over ellipsoidal shell), in the right column methods E1-E3 (integration over enclosed ellipsoidal volume).
	Cases where we removed the subhalos are marked with a --, cases where they remained by a +.
	It is evident, that it is essential to remove the subhalos in order to calculate the local shape correctly.}
	\label{fig:shapesubhalo}
\end{figure*}

Now we turn to a study of halos in cosmological structure formation simulations.
In these halos, in addition to the change of the axis ratios and the orientation of the principal axes as a function of distance, we also have subhalos.

The data are from a cosmological structure formation simulation, where we simulated several objects that will end up as Milky Way-sized objects at redshift $z=0$.
The simulations were run with the latest version of the gas dynamics and $N$-body adaptive refinement tree (ART) code \citep{1997ApJS..111...73K,1999PhDT........25K,2002ApJ...571..563K,2008ApJ...672...19R}.
ART includes 3-dimensional radiative transfer of ultraviolet (UV) radiation from individual stellar particles using the optically thin variable Eddington tensor (OTVET) approximation \citep{2001NewA....6..437G}.
It includes a non-equilibrium chemical network of hydrogen (H\,\textsc{i}, H\,\textsc{ii} and H$_2$) and helium (He\,\textsc{i}, He\,\textsc{ii} and He\,\textsc{iii}) as well as non-equilibrium cooling and heating rates, which use the local abundances of atomic, molecular and ionic species as well as the local UV intensity \citep{2011ApJ...728...88G}.
All these properties are followed self-consistently during the course of a simulation.
An empirical model for the formation and shielding of molecular hydrogen on the interstellar dust allows for more realistic star formation recipes based on the local density of molecular hydrogen \citep{2011ApJ...728...88G}.
Also included in ART is metal enrichment and thermal feedback due to the Type II and Type Ia supernovae \citep{2003ApJ...590L...1K} as well as stellar feedback \citep{2005ApJ...623..650K}.
Here, we use data at $z \approx 2$ from a simulation that includes cooling and star formation (simulation series A).
Further details are presented in an accompanying paper \citep{2011ZempBaryonicImpact}.

Figure \ref{fig:shapesubhalo} shows the shape of the total matter distribution of a massive halo at $z \approx 2$.
The distance is normalized by $r_\mathrm{200b}$ = 101 kpc, the radius that encloses a spherical volume such that the average enclosed density is 200 times the background density at that epoch.
The halo has a total mass $M_\mathrm{200b} = 1.13 \times 10^{12}~\Mo$ and contains $6.71 \times 10^6$ gas volume elements, $5.38 \times 10^6$ dark matter and $1.39 \times 10^6$ star particles within $r_\mathrm{200b}$.
All variants of the methods are shown with and without the subhalos from the resolution scale (0.003 $r_\mathrm{200b}$) up to 2 $r_\mathrm{200b}$.

Subhalos are removed by cutting out a spherical hole around the subhalo center with radius $r_\mathrm{trunc}$.
The spherical mass density profile of subhalos typically shows an uprise at large distances from their center due to the host halo.
The location where the minimum mass density is reached defines the truncation radius $r_\mathrm{trunc}$.
We investigated under controlled conditions the effects of cutting out holes of typical sizes of massive subhalos at different distances from the host halo center.
The deviations for the measured axis ratios at the location of the subhalo can be a few percent for methods S1 and E1 when compared to the smooth case.
Alternatively, one could only remove particles bound to subhalos \citep{2011ApJ...734...93L}.

For the halo shown in Figure \ref{fig:shapesubhalo}, the most massive subhalo has a mass of $9.02 \times 10^{9}~\Mo$, $r_\mathrm{trunc}$ = 2.77 kpc and is located at a distance of 10.5 kpc $\approx 0.1~r_\mathrm{200b}$ from the host halo center.
The total mass in all subhalos in this case is 3.71\% and most of the subhalos are located in the outer region of the halo.

Generally, the presence of massive subhalos leads to spikes in the axis ratios $b/a$ and $c/a$ when using an ellipsoidal shell as an integration volume (methods S1-S3).
The subhalos bias the measured axis ratios drastically at locations where they constitute a significant fraction of the total mass in the ellipsoidal shell.
These spikes are visible for all weight functions - most pronounced if no weighting or $w(\vec{r}) = r_\mathrm{ell}^{-2}$ is used and least pronounced for $w(\vec{r}) = r^{-2}$.
This effect is still present, though weaker, when integrating over the enclosed ellipsoidal volume without any weighting (method E1).
Often it is claimed in the literature that using the weights $r^{-2}$ or $r_\mathrm{ell}^{-2}$ in the shape tensor reduces the influence of subhalos on the shape determination.
This is true only if an enclosed integration volume is used (methods E2 and E3).

Again, there is nearly no difference between methods S1--, S1+, S3-- and S3+ if there are only few or no subhalos present at that distance (i.e. in the inner region).
If we integrate over the whole enclosed ellipsoidal volume, then the inclusion of the weighting by $r_\mathrm{ell}^{-2}$ smoothes out the detailed shape features.
Worse is using the weight $r^{-2}$ which leads again to a systematic shift of axis ratios towards larger values in our case in addition to the smoothing already observed for the $r_\mathrm{ell}^{-2}$ weighting, i.e. shapes are determined as rounder than they actually are.

\begin{figure}
	\includegraphics[width=\columnwidth]{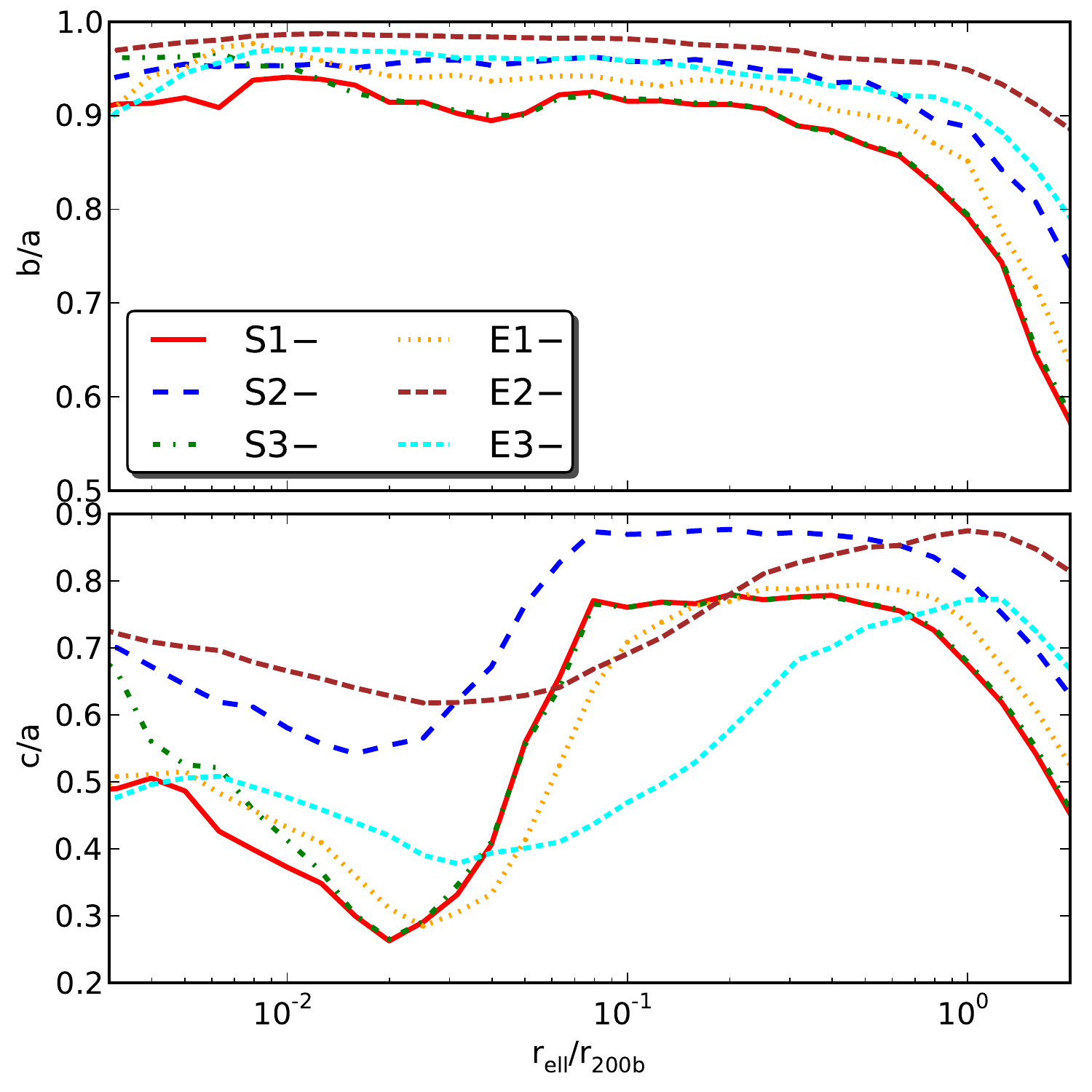}
	\caption{Measured median axis ratios $b/a$ (top panel) and $c/a$ (bottom panel) as a function of distance for the total matter distribution of the 16 halos at $z \approx 2$ for all the methods without the subhalos.}
	\label{fig:shapemedian}
\end{figure}

The median shape of the 16 most massive halos at $z \approx 2$ for all methods without the subhalos is shown in Figure \ref{fig:shapemedian}. 
Taking the median is motivated by the similarity of our selected halos \citep[within a factor of 10 in mass, see also][]{2011ZempBaryonicImpact}.
As already observed before, the methods S1-- and S3-- are nearly identical.
The methods where we integrate over an enclosed ellipsoidal volume (E1-E3) are naturally smoother than when integrating over an ellipsoidal shell volume (S1-S3).
As a consequence, the local shapes do not react as fast to shape changes in distance as in methods S1-S3, as seen for example for $c/a$.
There is a lag in distance when compared with shapes determined by methods that use an ellipsoidal shell as integration volume.
This is also visible for our single halo in Figure \ref{fig:shapesubhalo}.
For methods S2 and E2, the bias towards rounder shapes can be around 0.1--0.3 for both axis ratios.

\section{Discussion}

A widespread method used in the literature is method E3 \citep[e.g.][]{1991ApJ...378..496D,2006MNRAS.367.1781A,2007ApJ...671.1135K}.
By using an enclosed integration volume, this method picks up information from the inner regions that can have different shapes and orientation.
If one is interested in the local shape, then we find that method S1 is clearly a better choice than method E3.

Method E1 \citep[e.g.][]{1991ApJ...368..325K} is doing relatively well compared to its differential version S1.
This is due to the fact that the contribution in the shape tensor (Equation (\ref{eq:shapetensor})) is dominated by particles or volume elements with the largest distance from the center.
This method also shows systematic shifts (see for example Figure \ref{fig:shapemedian}) and smoothing when compared to method S1.
Therefore, the differential version S1 should be preferred over the E1 method that uses the enclosed ellipsoidal volume.

Unfortunately, methods S1 or S3 are not yet in widespread use in the literature.
\cite{2004ApJ...611L..73K} used method S3 and also found that using the enclosed volume is sensitive to the distribution of particles in the enclosed region.
Unfortunately, they did not present the details of the tests in their work.
\cite{2008ApJ...681.1076D} and \cite{2011ApJ...734...93L} are also advocating method S1.
While \cite{2008ApJ...681.1076D} do not further motivate their choice, \cite{2011ApJ...734...93L} found from visual comparison that using a differential method in 2 dimensions gives reliable ellipsoidal fits to X-ray isophotes.

\section{Summary}

We have critically examined different methods for determining the local shape of matter distributions as a function of distance.
Using the weights $r^{-2}$ or $r_\mathrm{ell}^{-2}$ in the shape tensor (Equation (\ref{eq:shapetensorgeneral})) does not cure the problem arising due to the presence of subhalos.
In contrary, it can lead to a systematic bias for the measured axis ratios even in smooth cases (Section \ref{sec:cc}).
We think it is better to remove the cause of the problem (i.e. the subhalos) than to fight the symptoms with weight factors that make the physical meaning of the shape tensor unclear.
Also when integrating over the whole enclosed ellipsoidal volume, features get smoothed out and shape changes are lagging behind in distance.

Therefore, our recommended method for measuring \textit{local} shapes is removing the subhalos, using ellipsoidal shells as the integration volume and determining the shape through an iteration method as described in Section \ref{sec:methods} that uses the shape tensor as defined in Equation (\ref{eq:shapetensor}), i.e. without any weight factors.

In some cases one is interested to characterize the shape of an object with just one number, i.e. one is not interested in the internal structure and the local shape as a function of distance.
Also, if the object is not well resolved (typically less than $\mathcal{O}(10^4)$ particles/volume elements), calculating the local shape can be problematic.
As a good practice, we recommend to have at least a few thousand particles in a bin when using ellipsoidal shells as integration volume.
Therefore, if the internal structure is not of interest or can not be properly resolved, we advocate method E1 since this shows the least bias among the tested methods that use the enclosed ellipsoidal volume.

\acknowledgements
\lastpagefootnotes

It is a pleasure to thank Jeremy Bailin, J{\"u}rg Diemand, Alexander Knebe and Mike Kuhlen for stimulating discussions and feedback on a draft of this paper.
MZ, OYG, NYG, and AVK are supported in part by NSF grant AST-0708087.
The simulations and analysis in this work have been performed on the Joint Fermilab-KICP Supercomputing Cluster (supported by grants from Fermilab, Kavli Institute for Cosmological Physics, and the University of Chicago) and the Flux cluster at the Center for Advanced Computing at the University of Michigan.
This research has made use of NASA's Astrophysics Data System (ADS), the arXiv.org preprint server, the visualization tool VisIt and the Python plotting library Matplotlib.

\bibliography{RDB_S}

\end{document}